\documentclass[debug,overfull,info]{epl}
\usepackage{amsmath}
\usepackage{epsf}
\title{
Iterated random walk
}
\author{L. Turban}
\institute{ 
  Laboratoire de Physique des Mat\'eriaux, UMR CNRS 7556, 
  Universit\'e Henri Poincar\'e (Nancy 1) 
  - BP 239, 54506 Vand\oe uvre l\`es Nancy Cedex, France
}

\begin{document}

\maketitle

\begin{abstract}
The iterated random walk is a random process in which a random walker moves on a 
one-dimensional random walk which is itself taking place on a one-dimensional 
random walk, and so on. This process is investigated in the continuum limit using 
the method of moments. When the number of iterations $n\to\infty$, a 
time-independent asymptotic density is obtained. 
It has a simple symmetric exponential form which is stable against the 
modification of a finite number of iterations. When $n$ is large, the deviation 
from the stationary density is exponentially small in $n$. The continuum results 
are compared to Monte Carlo data for the discrete iterated random walk. 
\end{abstract}

\section{Introduction}
When a walker moves at random along a straight support, its typical 
displacement at time $t$ grows like $l\sim t^{1/2}$. If the support 
of the walk is itself a random walk (RW),  
at time $t$ the walker is typically located at the $l$th step of the walk 
on which it is moving. Thus the mean-square displacement of 
the walker behaves as $\langle X^2(t)\rangle\sim l\sim 
t^{1/2}$ and his actual typical displacement is reduced, growing as
$t^{1/4}$ instead of $t^{1/2}$ for the conventional RW.  

This model of a RW on a RW has been studied for a one-dimensional 
support~\cite{kehr82} and also in higher dimensions~\cite{drager95,rabinovich96}. 
It has been used to discuss the diffusion of the stored length along a polymer 
chain entangled in a network~\cite{degennes71,degennes81,edwards85a} and the 
Brownian motion of charged particles in a turbulent plasma~\cite{edwards85b}. In 
this latter case, the particles are constrained to diffuse along the random 
magnetic field lines in the limit $B\to\infty$. 

\begin{figure}
\onefigure[width=7cm]{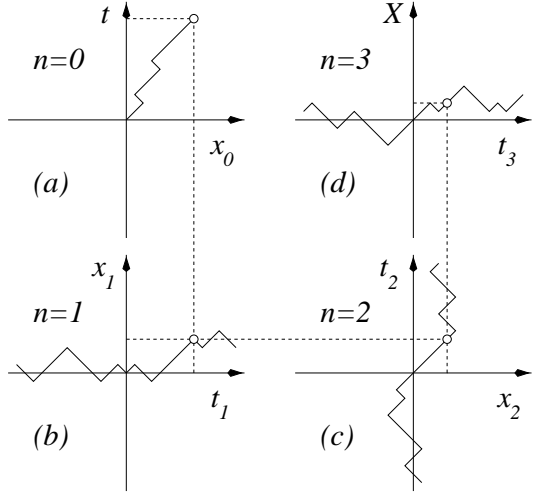}
\caption{Construction of the position $X$ of the random walker at time $t$ for 
the IRW: The walker in~(a) starts at $x_0(0)=X(0)=0$ and moves at random along 
the sequence of steps of the RW in~(b). Thus the coordinate $x_0(t)$ is also the 
time coordinate $t_1$ for the support in~(b). In the same way, $x_1$ gives the 
number of steps $t_2$ performed by the RW in~(c) on which the RW in~(b) takes 
place. Finally, at $t_3=x_2$, the last RW in~(d) is located at $X$, which gives 
the actual position $X(t)$ of the random walker.}
\label{f.1}
\end{figure}

In this work we study the properties of the one-dimensional iterated random 
walk (IRW), {\it i.e.}, the RW on a RW which is itself 
a RW on a RW, and so on. The first three steps of this 
iteration process are shown in fig.~\ref{f.1}. Working in the continuum 
(Gaussian) limit, we first study the moments $\langle X^p(t)\rangle$ of 
the probability density $P^{(n)}(X,t)$ to find the walker at $(X,t)$ after the 
$n$th iteration. In the next section, we examine the limit $n\to\infty$ for 
which the asymptotic probability density $P^{(\infty)}(X,t)$ can be obtained 
explicitly. We also look at the way this asymptotic density is approached when 
$n$ is large. Then the continuum result is compared to Monte Carlo data for the 
discrete IRW. The universality of the asymptotic probability 
density and possible generalizations are discussed in the final section.   

\section{Moments of the iterated probability density}

We assume that our random walker starts from $X=0$ at $t=0$. All other 
one-dimensional RWs, on which he is moving, extend in time from 
$-\infty$ to $+\infty$ and have a common origin. In the 
continuum limit, to the symmetric RW is associated the Gaussian density 
\begin{equation}
p(x,t)=\frac{1}{\sqrt{2\pi t}}\exp\left(-\frac{x^2}{2t}\right)\,,
\qquad t>0\,,
\label{e-gauss1}
\end{equation}
and the probabilty density to find the walker at $(X,t)$ after $n+1$ 
iterations is obtained by integrating the product of the successive probability 
densities $p(x_i,|t_i|)$ with $t_i=x_{i-1}$, $t_0=t$ for the random walker and 
$x_{n+1}=X$ for the last support, over all the intermediate coordinates $x_i$ for 
$i=0$ to $i=n$ (see fig.~\ref{f.1}):  
\begin{equation}
P^{(n+1)}(X,t)=\int_{-\infty}^{+\infty}\!\!\!\!\upd x_n\, p(X,|x_n|)
\int_{-\infty}^{+\infty}\!\!\!\!\upd x_{n-1}\, p(x_n,|x_{n-1}|)\cdots
\int_{-\infty}^{+\infty}\!\!\!\!\upd x_0\, p(x_1,|x_0|)p(x_0,t)\,.
\label{e-iter1}
\end{equation}
As a consequence, the iterated probability density satisfies the recursion 
relation:
\begin{equation}
P^{(n+1)}(X,t)=\int_{-\infty}^{+\infty}\!\!\!\!\upd x\, p(X,|x|)P^{(n)}(x,t)\,.
\label{e-recur1}
\end{equation}
In the following we leave the initial probability density 
$P^{(0)}(x,t)$ unspecified. Since the Gaussian density $p(X,|x|)$ 
in~(\ref{e-recur1}) is even in $X$, the same is true of all the iterated 
densities $P^{(n)}(x,t)$, $n>0$. Their odd moments vanish while their even 
moments are given by
\begin{equation}
M_{2p}^{(n)}(t)=\int_{-\infty}^{+\infty}\!\!\!\!\upd X\, X^{2p} 
P^{(n)}(X,t)={\cal M}_{2p}^{(n)}(t)\,,\quad 
{\cal M}_{p}^{(n)}(t)=\int_{-\infty}^{+\infty}\!\!\!\!\upd X\, |X|^p 
P^{(n)}(X,t)\,,
\label{e-mom1}
\end{equation}
where the moments ${\cal M}_{p}^{(n)}$ of $|X|$ have been introduced.
According to~(\ref{e-recur1}), these moments may be written as 
\begin{equation}
{\cal M}_{p}^{(n+1)}(t)=\!\int_{-\infty}^{+\infty}\!\!\!\!\upd x\, P^{(n)}(x,t)
\int_{-\infty}^{+\infty}\!\!\!\!\upd X\, |X|^p p(X,|x|)
=2^{p/2}\frac{\Gamma\left(\frac{p}{2}+\frac{1}{2}\right)}
{\Gamma\left(\frac{1}{2}\right)}\!
\int_{-\infty}^{+\infty}\!\!\!\!\upd x\, |x|^{p/2} P^{(n)}(x,t)\,,
\label{e-smom1}
\end{equation}
where the last relation follows from the form of $\langle|X|^p\rangle$ for the 
Gaussion density,
\begin{equation}
\int_{-\infty}^{+\infty}\!\!\!\!\upd X\, |X|^p\, 
\frac{\exp\left(-\frac{X^2}{2t}\right)}{\sqrt{2\pi t}}
=\frac{\Gamma\left(\frac{p}{2}+\frac{1}{2}\right)}
{\Gamma\left(\frac{1}{2}\right)}\,(2t)^{p/2}\,.
\label{e-smom2}
\end{equation}
Equation~(\ref{e-smom1}) leads to the recursion relation
\begin{equation}
{\cal M}_{p}^{(n+1)}(t)=2^{p/2}\frac{\Gamma\left(\frac{p}{2}
+\frac{1}{2}\right)}
{\Gamma\left(\frac{1}{2}\right)}\,{\cal M}_{p/2}^{(n)}(t)\,,
\label{e-recur2}
\end{equation}
so that
\begin{equation}
{\cal M}_{p}^{(n)}(t)={\cal M}_{p/2^n}^{(0)}(t)
\prod_{k=1}^n 2^{p/2^k}\frac{\Gamma\left(\frac{p}{2^k}+\frac{1}{2}\right)}
{\Gamma\left(\frac{1}{2}\right)}\,.
\label{e-smom3}
\end{equation}
Making use of the identity~\cite{hansen75}
\begin{equation}
\Pi(z)=\prod_{k=1}^n
\frac{\Gamma\left(\frac{z}{2^k}+\frac{1}{2}\right)}
{\Gamma\left(\frac{1}{2}\right)}=\frac{\Gamma(z+1)}
{2^{2z(1-1/2^n)}
\Gamma\left(\frac{z}{2^n}+1\right)}\,,
\label{e-prod}
\end{equation}
the moments of the iterated probability density may be written as:
\begin{equation}
M_{2p}^{(n)}(t)=\frac{(2p)!}{2^{2p(1-1/2^n)}
\Gamma\left(\frac{2p}{2^n}+1\right)}\, {\cal M}_{2p/2^n}^{(0)}(t)\,,
\qquad 
M_{2p+1}^{(n)}(t)=0\,.
\label{e-mom2}
\end{equation}

\section{Asymptotic probability density and how it is approached}
Since the walker and all the underlying RWs have a common origin in 
space and time,
\begin{equation}
P^{(n)}(X,0)=\delta(X)\,.
\label{e-origin}
\end{equation}
When $t>0$ and $n\to\infty$, the even moments in~(\ref{e-mom2}) considerably 
simplify, 
\begin{equation}
M_{2p}^{(\infty)}(t>0)=\frac{(2p)!}{4^{p}}\,,
\label{e-mom3}
\end{equation}
and become time-independent since ${\cal M}_0^{(0)}(t)=1$ for any initial 
density function. The Fourier transform of the asymptotic probability density is 
related to its moments through a Taylor expansion:
\begin{eqnarray}
{\cal P}^{(\infty)}(k,t>0)&=&\int_{-\infty}^{+\infty}\!\!\!\!\upd X\,  
{\mathrm e}^{{\mathrm i} kX} P^{(\infty)}(x,t>0)
=\sum_{q=0}^{\infty}\frac{({\mathrm i} k)^q}{q!}M_q^{(\infty)}(t>0)\nonumber\\
&=&\sum_{p=0}^{\infty}\left(-\frac{k^2}{4}\right)^p=\frac{4}{k^2+4}\,.
\label{e-tf}
\end{eqnarray}
The inverse Fourier transform,
\begin{equation}
P^{(\infty)}(X,t>0)=\frac{1}{2\pi}\int_{-\infty}^{+\infty}\!\!\!\!\upd k\,  
{\mathrm e}^{-{\mathrm i} kX}{\cal P}^{(\infty)}(k,t>0)
=\frac{2}{\pi}\int_{-\infty}^{+\infty}\!\!\!\!\upd k\,
\frac{{\mathrm e}^{-{\mathrm i} kX}}{k^2+4}\,,
\label{e-tfi}
\end{equation}
is easily evaluated using the method of residues. Finally, the asymtotic 
density takes the simple form
\begin{equation}
P^{(\infty)}(X,t)=\left\{
\begin{array}{ll}
\delta(X)&(t=0)\\
\noalign{\vspace{3pt}}
{\mathrm e}^{-2|X|}&(t>0)
\end{array}
\right.\,,
\label{e-pasym}
\end{equation}
which is {\it time-independent} as soon as $t>0$ and {\it universal}, {\it 
i.e.}, independent of the form of the initial density $P^{(0)}$. This asymptotic 
density is invariant under the Gaussian transformation~(\ref{e-recur1}).

Let us now examine how this stationary density is approached when $n$ is 
large. For this, the initial density $P^{(0)}$ must be specified. Coming back to 
the random-walk problem, we use the Gaussian density for which
\begin{equation}
{\cal M}_{2p/2^n}^{(0)}(t)=
\frac{\Gamma\left(\frac{p}{2^n}+\frac{1}{2}\right)}
{\Gamma\left(\frac{1}{2}\right)}\,(2t)^{p/2^n}
\label{e-mom4}
\end{equation}
according to~(\ref{e-smom2}). The even moments in~(\ref{e-mom2}) take the 
following form:
\begin{equation}
M_{2p}^{(n)}(t)=\frac{(2p)!}{4^p}
\frac{
2^{2p/2^n}\Gamma\left(\frac{p}{2^n}+\frac{1}{2}\right)
}
{
\Gamma\left(\frac{2p}{2^n}+1\right)\Gamma\left(\frac{1}{2}\right)
}
\,(2t)^{p/2^n}\,.
\label{e-mom5}
\end{equation}
Using the duplication formula~\cite{abramowitz72},
\begin{equation}
\frac{\Gamma(2z)}{\Gamma(z)}=2^{2z-1}
\frac{\Gamma\left(z+\frac{1}{2}\right)}{\Gamma\left(\frac{1}{2}\right)}\,,
\label{e-dupli}
\end{equation}
this expression can be rewritten as:
\begin{equation}
M_{2p}^{(n)}(t)=\frac{(2p)!}{4^p}
\frac{(2t)^{p/2^n}}{\Gamma\left(1+\frac{p}{2^n}\right)}\,.
\label{e-mom6}
\end{equation}
Thus $M_{2p}^{(n)}(0)=\delta_{p,0}$ in agreement with~(\ref{e-origin}) and a
length scales as~$t^{1/2^{n+1}}$ after $n$ iterations.

When $n\gg1$, a first-order expansion in powers of $2^{-n}$ gives
\begin{eqnarray}
(2t)^{p/2^n}
&=&1+\ln(2t)\frac{p}{2^n}+\mathord{\mathrm{O}}(4^{-n})\nonumber\\
\Gamma\left(1+\frac{p}{2^n}\right)
&=&1+\Psi(1)\frac{p}{2^n}+\mathord{\mathrm{O}}(4^{-n})
=1-\gamma\frac{p}{2^n}+\mathord{\mathrm{O}}(4^{-n})\nonumber\\
M_{2p}^{(n)}(t)&=&\frac{(2p)!}{4^p}
\left\{1+[\ln(2t)+\gamma]\frac{p}{2^n}\right\}+\mathord{\mathrm{O}}(4^{-n})\,,
\label{e-expan}
\end{eqnarray}
where $\Psi(z)=\Gamma'(z)/\Gamma(z)$ is the digamma function and 
$\gamma=.577\,215\,664\ldots$ is Euler's constant.  
Thus, the leading correction to the Fourier transform of the asymptotic density 
reads
\begin{equation}
\delta{\cal P}^{(n)}(k,t>0)=\frac{\ln(2t)+\gamma}{2^n}\sum_{p=1}^\infty
p\left(-\frac{k^2}{4}\right)^p
=-\frac{\ln(2t)+\gamma}{2^n}\frac{4k^2}{(k^2+4)^2}
\label{e-corrk}
\end{equation}
and the inverse Fourier transform leads to:
\begin{equation}
\delta P^{(n)}(X,t>0)=\frac{\ln(2t)+\gamma}{2^n}\,\left(|X|-1\right)\,{\mathrm 
e}^{-2|X|}\,.
\label{e-corrx}
\end{equation}
The deviation from the asymptotic density is exponentially small in $n$, a 
behaviour which is independent of the initial density function according 
to~(\ref{e-mom2}). It has a weak (logarithmic) time dependence for the Gaussian 
density. 

\begin{figure}
\onefigure[width=7cm]{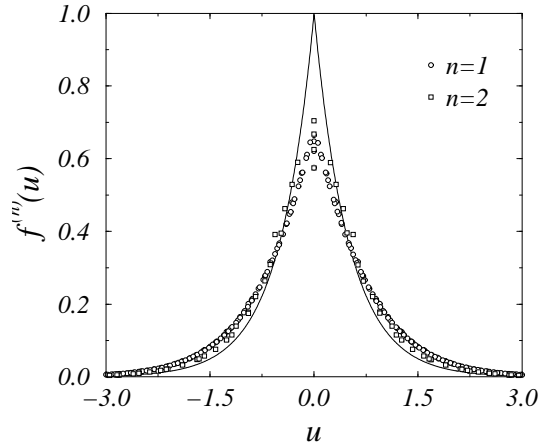}
\caption{Scaling function for the iterated probability distributions of the 
discrete RW compared to the stationary density (solid line) obtained in 
the continuum limit. The histograms of the probability distributions 
have been obtained through Monte Carlo simulations for $t=10^2$ to $10^5$ and 
$t+1$, working with $10^7$ samples. The statistical fluctuations are much 
smaller than the symbols as indicated by the symmetry of the data.}
\label{f.2}
\end{figure}

\section{Scaling behaviour and discrete IRW}
The Gaussian density in~(\ref{e-gauss1}) can be written under the more general 
scaling form
\begin{equation}
p(x,t)=t^{-1/z}f\left(\frac{x}{t^{1/z}}\right)\,,
\label{e-scal1}
\end{equation}
where the dynamical exponent $z$, which is equal to 2 for the Gaussian density, 
governs the transformation of the time, $t'=t/b^z$, under a change of the length 
scale, $x'=x/b$. The scaling behaviour of the iterated density can be obtained 
by induction. Assuming that 
\begin{equation}
P^{(n)}(X,t)=t^{-1/z_n}
f^{(n)}\left(\frac{X}{t^{1/z_n}}\right)\,,
\label{e-scal2}
\end{equation}
which is true with $z_0=z$ and $n=0$ for the initial density, and making use 
of~(\ref{e-recur1}), we obtain
\begin{eqnarray}
P^{(n+1)}(X,t)&=&t^{-1/z_n}
\int_{-\infty}^{+\infty}\!\!\!\!\upd x\, |x|^{-1/z}
f\!\left(\frac{X}{|x|^{1/z}}\right) 
f^{(n)}\left(\frac{x}{t^{1/z_n}}\right)\nonumber\\
&=&t^{-1/zz_n}
\int_{-\infty}^{+\infty}\!\!\!\!\upd v\,
f\!\left(\frac{X}{|v|^{1/z}t^{1/zz_n}}\right)\frac{f^{(n)}(v)}{|v|^{1/z}}\,,
\label{e-recur3}
\end{eqnarray}
so that $P^{(n+1)}$ scales like $P^{(n)}$ in~(\ref{e-scal2}). The dynamical 
exponent evolves according to $z_{n}=zz_{n-1}=z^{n+1}$ and the scaling function 
satisfies the recursion relation
\begin{equation}
f^{(n)}(u)=\int_{-\infty}^{+\infty}\!\!\!\!\upd v\,
f\!\left(\frac{u}{|v|^{1/z}}\right)\frac{f^{(n-1)}(v)}{|v|^{1/z}}\,
\qquad u=\frac{X}{t^{1/z^{n+1}}}\,.
\label{e-recur4}
\end{equation}
Thus, with $z>1$, the asymptotic density $P^{(\infty)}(X,t)$ is time-independent 
and reduces to its scaling function $f^{(\infty)}(X)$.

We now make use of this scaling behaviour to analyse Monte Carlo data for the 
discrete IRW. The continuum result is expected to give a 
good description of these data for not too small values of $X$ and $t$. The 
histograms for the discrete probability distributions have been obtained as 
follows: First a RW with $t$ steps is generated, leading the walker to 
$x_0$ (see~fig.~\ref{f.1}). Then a second walk with $|x_0|$ steps, corresponding 
to the first support, is constructed. When it ends at $x_1$ the value of 
$N^{(1)}(x_1)$, giving the number of walks ending at $x_1$ at the first 
iteration, is updated and the same process is repeated for the following 
iterations. $N_s=10^7$ samples have been generated in this way for 
$t=10^2,10^3,10^4,10^5$. Since all the $x_n$ have the parity of $t$, the 
simulations were repeated for $t+1$ in order to double the number of points 
along the $X$ axis. The properly normalized histograms for a comparison with the 
density functions are obtained by dividing $N^{(n)}$ by $2N_s$.

The scaling functions for the discrete iterated walk are compared to the 
invariant density~(\ref{e-pasym}) in fig.~\ref{f.2} . A good data collapse 
is obtained for each value of $n$. As expected, there is a rapid 
convergence to the asymptotic density. The convergence is from above when $|u|$ 
is greater than a value close to 1, in agreement with~(\ref{e-corrx}). 

\section{Final remarks} We have seen that the Gaussian iteration process leads 
to an invariant probability density~(\ref{e-pasym}) which is independent of the 
initial density $P^{(0)}$. Actually, the universality of the invariant density 
is more extended since $P^{(0)}$ may be itself considered as resulting from a 
finite number of iterations involving arbitrary densities $p_k(x,t)$ $(k=0,m)$. 
The initial density being now $p_0(x_0,t)$, we have:
\begin{eqnarray}
P^{(0)}(x,t)&=&\int_{-\infty}^{+\infty}\!\!\!\!\upd x_{m-1}\, p_m(x,|x_{m-1}|)
\int_{-\infty}^{+\infty}\!\!\!\!\upd x_{m-2}\, p_{m-1}(x_{m-1},|x_{m-2}|)\cdots
\nonumber\\
&&\ \ \ \ \cdots\int_{-\infty}^{+\infty}\!\!\!\!\upd x_0\, 
p_1(x_1,|x_0|)p_0(x_0,t)\,.
\label{e-iter2}
\end{eqnarray}
After an infinite number of Gaussian iterations this density will evolve, as 
before, to the universal time-independent invariant density. The arbitrary 
densities may be, for example, shifted Gaussian densities
\begin{equation}
p_k(x,t)=\frac{1}{\sqrt{2\pi t}}\exp\left[-\frac{(x-v_kt)^2}{2t}\right]
\label{e-gauss2}
\end{equation}
corresponding to RWs with arbitrary drifts $v_k$.

When, in the iteration process, the Gaussian density is replaced by a density  
which is even in $x$, scales like in~(\ref{e-scal1}) and has finite moments, the 
nonvanishing even moments of the iterated density are easily obtained in the 
same way as above. They involve a product similar to that appearing 
in~(\ref{e-smom3}) and are time-independent when $n\to\infty$. Unfortunately, we 
were unable to find another instance where these moments could be simplified as 
in~(\ref{e-mom2}), allowing for an explicit calculation of the time-independent 
asymptotic density.


\begin{thebibliography}{0}

\bibitem{kehr82}
  \Name{Kehr K. W. \and Kutner R.}
  \REVIEW{Physica A}{110}{1982}{535}.

\bibitem{drager95}
  \Name{Dr\"{a}ger J., Russ S., \and Bunde A.}
  \REVIEW{Europhys. Lett.}{31}{1995}{425}.

\bibitem{rabinovich96}
  \Name{Rabinovich S., Roman H. E., Havlin S. \and Bunde A.}
  \REVIEW{Phys. Rev. E}{54}{1996}{3606}.

\bibitem{degennes71}
  \Name{de Gennes P. G.}
  \REVIEW{J. Chem. Phys.}{55}{1971}{572}.

\bibitem{degennes81}
  \Name{de Gennes P. G.}
  \REVIEW{J. Phys. (Paris)}{42}{1981}{735}.

\bibitem{edwards85a}
  \Name{Edwards S. F.}
  \REVIEW{Br. Polym. J.}{17}{1985}{122}.

\bibitem{edwards85b}
  \Name{Edwards S. F.}
  \REVIEW{Phil. Mag. B}{52}{1985}{573}.

\bibitem{hansen75}
  \Name{Hansen E. R.}
  \Book{A Table of Series and Products}
  \Publ{Prentice-Hall, Englewood Cliffs, N. J.}
  \Year{1975}
  \Page{496}. This identity is easily obtained by building the ratio 
  $\Pi(2z)/\Pi(z)$ and making use of the duplication 
  formula~(\protect\ref{e-dupli}).
  
\bibitem{abramowitz72}
  \Name{Abramowitz M. \and Stegun I. A.}
  \Book{Handbook of Mathematical Functions}
  \Publ{Dover Publications, New York}
  \Year{1972}
  \Page{256}.

\end{thebibliography}
\end{document}